\documentclass[conference]{./IEEEtran}
\usepackage{graphicx}

\usepackage{tikz}

\usepackage{xcolor}

\usepackage{amsmath}
\usepackage{mathrsfs}

\usepackage{amsthm}
\theoremstyle{definition}

\usepackage{amssymb}

\usepackage{algorithm}
\usepackage{algorithmic}

\def\varx{\mathbf{x}}
\def\varb{\mathbf{b}}
\def\vard{\mathbf{d}}

\def\mycolor{black}

\def\whichyear{2021}
\usepackage{tikz}
\usepackage{textcomp}
\usepackage{lipsum}
\newcommand\copyrighttext{%
  \footnotesize \textcopyright \whichyear \hspace{0.2em} IEEE. Personal use is permitted. This is the author's version of an article that has been published in this journal. Changes were made to this version by the publisher prior to publication.}
\newcommand\copyrightnotice{%
\begin{tikzpicture}[remember picture,overlay]
\node[anchor=south,yshift=10pt] at (current page.south) {\fbox{\parbox{\dimexpr\textwidth-\fboxsep-\fboxrule\relax}{\copyrighttext}}};
\end{tikzpicture}%
}

\begin{document}

\title{Dynamic Bandwidth Allocation for PON Slicing with Performance-Guaranteed Online Convex Optimization}
\author{
    \IEEEauthorblockN{Genya Ishigaki, Siddartha Devic, Riti Gour, and Jason P. Jue}
    \IEEEauthorblockA{
        Department of Computer Science, The University of Texas at Dallas, Richardson, Texas 75080, USA\\
        Email: \{gishigaki, sid.devic, rgour, jjue\}@utdallas.edu
    }
}

\maketitle

\copyrightnotice

\IEEEpeerreviewmaketitle

\begin{abstract}
The emergence of diverse network applications demands more flexible and responsive resource allocation for networks.
Network slicing is a key enabling technology that provides each network service with a tailored set of network resources to satisfy specific service requirements. The focus of this paper is the network slicing of access networks realized by Passive Optical Networks (PONs). 
This paper proposes a learning-based Dynamic Bandwidth Allocation (DBA) algorithm for PON access networks, considering slice-awareness, demand-responsiveness, and allocation fairness. Our online convex optimization-based algorithm learns the implicit traffic trend over time and determines the most robust window allocation that reduces the average latency.
Our simulation results indicate that the proposed algorithm reduces the average latency by prioritizing delay-sensitive and heavily-loaded ONUs while guaranteeing a minimal window allocation to all ONUs.
\end{abstract}



\section{Introduction}


%

Emerging 5G networks are being designed to support a diverse array of network applications. Enhanced Mobile Broadband (eMBB) targets up to 10 Gbit/s download speeds for mobile devices, while massive Machine Type Communications (mMTC) handles dense connections for IoT and M2M applications, and Ultra Reliable Low Latency Communications (URLLC) involves strict latency constraints \cite{8685766}. An approach for accommodating massive and diverse connections in 5G access networks is to implement network slicing in Passive Optical Networks (PONs) \cite{ntt_pon}. Optical access networks have attracted many industrial attentions, as they are expected to reduce operational expenditure (OPEX) \cite{6461187}. In particular, the current development of C-RAN technology for mobile networks accelerates the benefit of such optical access networks \cite{6897914, 8384342}. 

\begin{figure}[t]
    \centering
    \includegraphics[width=\columnwidth]{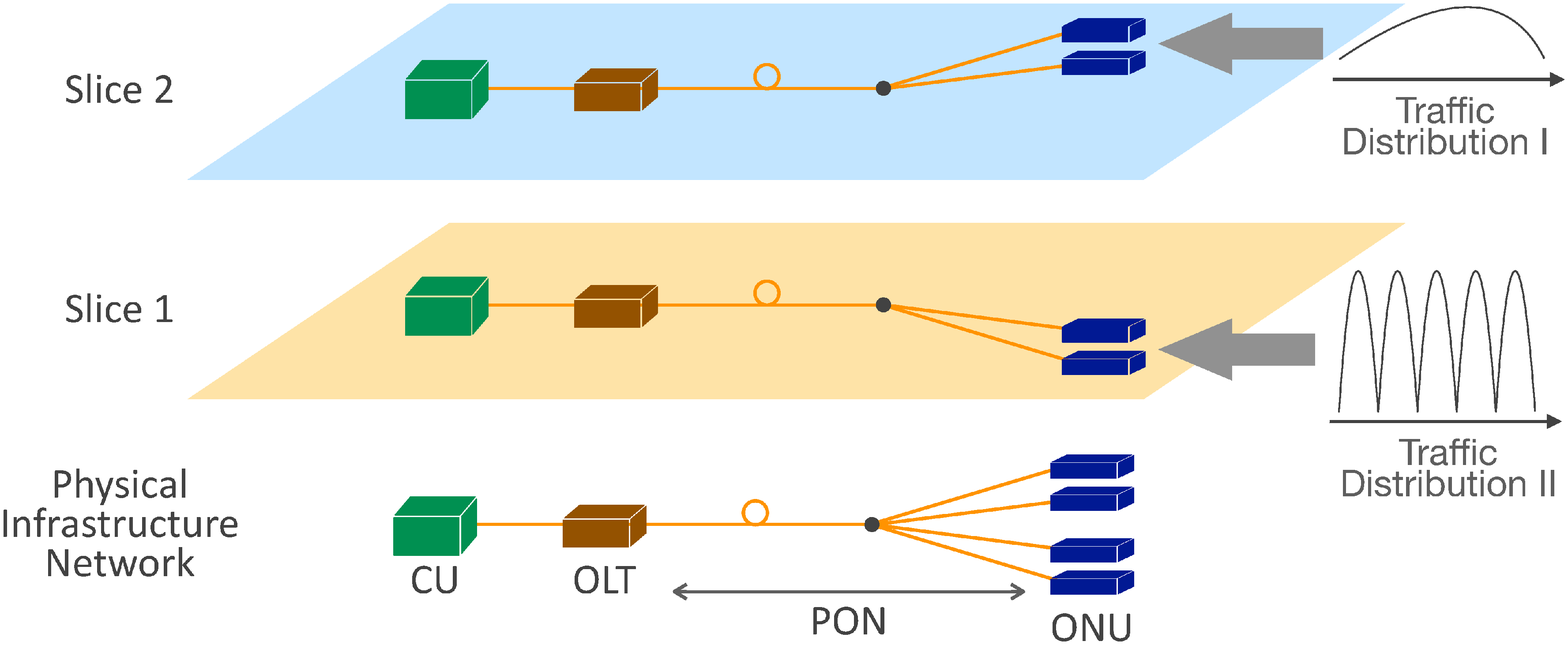} 
    \caption{PON Access Network Architecture: A physical network is sliced to support different application requirements.}
    \label{fig_arch}
\end{figure}

A PON consists of an Optical Line Terminal (OLT) and multiple Optical Network Units (ONUs) as depicted in Figure \ref{fig_arch}. The OLT and ONUs may be logically sliced so that one physical network can host multiple network infrastructure slices that are tailored for various types of applications. In this slicing scenario, additional consideration should be made to prioritize delay-critical network slices. For example, if a network slice hosts a URLLC network service, the ONUs on the slice should receive more bandwidth to meet the strict delay requirement on the order of $1$ ms for both upstream and downstream traffic \cite{8985528}.


Time Division Multiplexing (TDM) for the upstream traffic from ONUs to an OLT is facilitated by a \textit{Dynamic Bandwidth Allocation} (DBA) algorithm that adjusts time windows for each ONU to minimize the latency of upstream traffic \cite{4804388survey}. Conventional DBA algorithms periodically solve a static optimization problem to minimize the traffic latency by choosing an optimal bandwidth allocation given traffic information of the ONUs \cite{9003209,8765309,7937095,8647809}. The traffic information can be an estimation of the future traffic amount at the time when the ONU starts sending traffic to the OLT. The estimation is computed based on historical data, the amount of remaining data in queues, and/or neural network-based traffic prediction. Nevertheless, errors in such estimations may lead to suboptimal solutions for the bandwidth allocation problem. The suboptimality ironically originates from the fact that such estimation-based algorithms try to achieve the optimum window allocation in each static optimization problem, which is defined based on the estimated traffic demand. Since the allocation is optimum only for the estimated demand, the discrepancy between the estimation and the real traffic patterns is directly reflected in the performance of the algorithms. 

Therefore, we propose a performance-guaranteed online DBA algorithm based on Online Convex Optimization (OCO), which does not involve static optimizations based on an unreliable estimation of traffic patterns. An OCO algorithm generates a sequence of decisions, observing the actual past traffic information rather than an estimation of future states. The decision sequence will be adjusted along its runs to minimize the \textit{regret}, the loss incurred by diverting from a decision that is robust over the entire time horizon. Note that a truly robust decision can be defined only in retrospect. In the PON context, the OCO-based DBA algorithm decides a window allocation for the next time cycle based on the historical window sizes and \textit{actual} traffic information observed in the previous cycles, so that the resulting allocation becomes robust against the fluctuating traffic. In other words, the algorithm implicitly learns a potential distribution of future traffic and provides a safe window allocation that can accommodate any traffic dynamics that may happen under the distribution. Furthermore, the performance of the algorithm can be theoretically bounded in terms of regret \cite{shalev2011online}. Our algorithm is more promising than DBA algorithms based on traffic estimation, since its allocation would realize lower latency within a potential traffic range, instead of being optimal only at one estimated traffic amount. Our DBA problem is formulated in such a way as to ensure fairness among ONUs, which guarantees that each ONU receives a minimal size of time window even when it is not heavily loaded. This formulation could be seen as an extension of the work in \cite{7830276} to a dynamic setting. The simulation results show that our OCO-based DBA algorithm reduces the latency at all ONUs, demonstrating slice-awareness, demand-responsiveness, and allocation fairness.


\begin{figure}[t]
    \centering
    \includegraphics[width=\columnwidth]{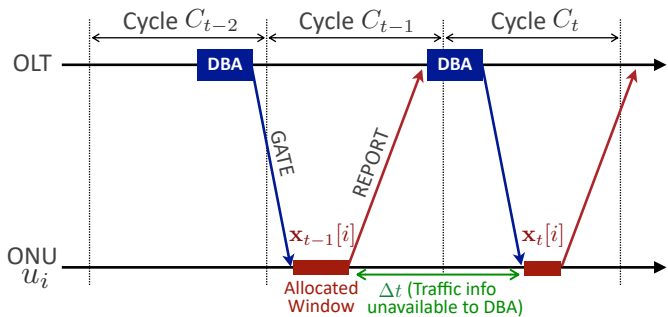} %
    \caption{{\color{\mycolor}Conventional} interactions between OLT and ONUs: The OLT decides a bandwidth allocation $\varx_t[i]$ to each ONU $u_i$ based on the demand request included in a REPORT message.}
    \label{fig_cycle}
\end{figure}

\section{PON Access Network Model}
A Passive Optical Network (PON) consists of an Optical Line Terminal (OLT) and Optical Network Units (ONUs). Since all the ONUs share one optical cable connecting them to the OLT, they need to be coordinated to share the bandwidth for upstream traffic. Time Division Multiplexing (TDM) PON realizes the sharing by allocating a time window to each ONU in each logical time cycle.

Figure \ref{fig_cycle} illustrates {\color{\mycolor}conventional interactions of an OLT and an ONU $u_i \in U$ in a PON.} An OLT has a Dynamic Bandwidth Allocation (DBA) algorithm that decides the allocation to ONUs. In EPON or 10G EPON standards, an ONU sends a REPORT message that includes a demand request for its next turn. The request is generated by an traffic estimation algorithm at ONUs, which may consider the amount of currently buffered traffic and the future traffic pattern. Receiving the request, the OLT decides the actual bandwidth allocation for the ONU and informs the ONU of the assigned time windows through a GATE message.

{\color{\mycolor}
However, it is pointed out that the DBA computation using the latest requests from ONUs causes an extra delay to grant the next TDM windows \cite{9046150}. Therefore, some works even propose DBA algorithms that do not use the REPORT messages. For example, Cooperative DBA algorithms (CO-DBAs) \cite{6886953} compute the future bandwidth allocation based on the mobile schedule information provided by a Base Band Unit (BBU), which is not an element of a PON itself, instead of waiting for the requests from ONUs. The allocation decision that does not rely on the ONU requests enables CO-DBAs to reduce the overall latency. 

Based on the discussion about the negative effect of request messages, we also consider a bandwidth allocation problem in a PON model that is slightly different from the conventional OLT/ONU interactions. Our model does not have allocation requests from ONUs, which cause the extra delay and could include unreliable statistics such as outdated queue state or estimated future traffic demands. The REPORT messages are repurposed so that they report the actual traffic amount, which implies the discrepancy between the actual window size needed and the actual allocation executed in the previous time cycle. In other words, GATE messages to grant next upstream windows are generated based on not the ONU requests but the allocation decision computed from the history of the actual traffic amount buffered at each ONU.
}

In this paper, we assume that an OLT divides the time horizon into multiple logical time cycles $\{C_t\}\ (t = 1, ..., T)$ and solves the bandwidth allocation for each time cycle. Note that each ONU receives its turn to send out upstream traffic within the time cycle. With the notation, the amount of bandwidth allocated to an ONU $u_i \in U$ at a cycle $C_t$ is denoted as $\varx_t[i]$.



\section{Problem Statement}
Our DBA problem is formulated to provide \textit{slice-aware}, \textit{demand-responsive}, and \textit{fair} bandwidth allocation for ONUs. The slice awareness indicates the capability to satisfy additional delay requirements imposed by some slices. The demand responsiveness is defined as the ability to adjust window sizes based on the amount of traffic loads at ONUs. Furthermore, the bandwidth allocation is fair when some ONUs with huge traffic do not dominate a cycle; i.e., each ONU receives an opportunity to send a minimal amount of its queued traffic within each cycle.

The concept of \textit{proportional fairness} {\color{\mycolor}\cite{879343}} is introduced to formulate the DBA problem incorporating the three properties. A vector $\varx^*$ is said to be $(\mathbf{w}, 1)$-proportionally fair if the proportions of value differences in any other allocations $\varx \in X$ sum to negative:
\begin{align}
\centering
\sum_{i} \mathbf{w}_{i} \frac{\varx[i]-\varx^{*}[i]}{\varx^{*}[i]} \leq 0\ (\forall \varx \in X). 
\end{align}
{\color{\mycolor} Here, the weight $\mathbf{w}_i$ indicates the relative importance of component $i$, which receives a resource quota.}

It is known that a maximizer $\varx^*$ for the objective function of the following form over a convex space $X = \{\varx\}$ satisfies the proportional fairness \cite{7830276}.
\begin{align}
    \sum_i \mathbf{w}_i\log \varx[i].    
    \label{pf_obj}
\end{align}

The intuition behind the relationship between the objective function in Eq. \eqref{pf_obj} and the fairness concept comes from the decreasing increment of logarithmic functions. Let us consider two points $x_1$ and $x_2$ such that $x_1 < x_2$. When evaluating the increment of a logarithmic function $h$ with the same increment $\delta$ from each point, $h(x_1 + \delta) > h(x_2 + \delta)$ always holds. 

In our context, the increment of our objective function, which represents the utility of a specific bandwidth allocation, becomes larger when providing more bandwidth to an ONU that is assigned a smaller bandwidth window. This property enables us to avoid the domination of a time cycle by a small number of ONUs.

The DBA problem is to determine a time partition $\varx_t = (\varx_t[i])_{i = 1, ..., |U|} $ of a cycle $C_t$, where each element in the partition corresponds to a window size allocated to an ONU $u_i \in U$. Let $C \in \mathbb{R}_+$ denote the fixed cycle duration where every ONU receives a fraction of the cycle. An ONU slice $S_j \subseteq U$ is defined as a subset of ONUs that are running on a network slice $j$, and the prioritization weight $p_j$ indicates the sensitivity of the slice $S_j$ to delay.

Assuming the availability of accurate information regarding the amount $\varb_t = (\varb_t[i])_{i = 1, ..., |U|}$ of data queued in every ONU $u_i$ at the starting time of its window in $C_t$, the optimization problem, which maximizes the total weighted utility of allocated windows, is defined as follows:
\begin{align}
\textrm{Maximize } f_t(\mathbf{x}_t) = &\sum_i \varb_t[i] p_{j: u_i \in S_j} \min\{\log (\varx_t[i] + 1),\notag\\ 
& \log (\varb_t[i] + 1)\}\\
&\textrm{subject to}\ \mathbf{x}_t^\top \mathbf{1}_N \leq C - \sum_i \vard[i],
\label{objective}
\end{align}
where $\vard$ is a vector of guard window sizes that prevent collisions when switching ONUs.

The min function prevents the utility function from increasing by overallocation. The maximum utility that an ONU $u_i$ can experience at a cycle $C_t$ must be bounded by the allocation that empties its entire buffer ($\varx_t[i] = \varb_t[i]$). When the allocation is less than the emptying allocation, the utility of the ONU should be represented as an increasing function over the allocation $\varx$. To represent these property, the objective cuts off the increase in the utility by taking a minimum between two logarithm values. 

Since the constraint in Eq. \eqref{objective} defines a scaled simplex of $\varx_t$, the decision space $X$ is a convex set. The objective function $f$ has the exact same form as the proportionally fair optimization discussed above, recognizing the pair of variables $\varb_t[i]$ and  $p_{j: u_i \in S_j}$ as a weight $\mathbf{w}_i$. 

In addition, the fairness is defined with respect to two weights $\varb$ and $p$ that indicate the load on ONUs and the prioritization weight of slices, respectively. Therefore, the maximization with these weights realizes the slice awareness and demand responsiveness, since ONUs with more traffic load and/or with more stringent delay requirements would receive more bandwidth allocations.


\section{Performance-Guaranteed Dynamic Bandwidth Allocation Based on OCO}
\subsection{OCO-based DBA Algorithm}
With the assumption of an oracle to report the accurate amount of queued data $\varb_t$, the defined DBA problem is easy to solve, as it has a concave objective function and a convex variable space. However, the practical difficulty of DBA problems arises from the fact that it is impossible to obtain the actual data amount $\varb_t$ at the starting time of the window allocated to each ONU. This is because the demand request in a REPORT message is computed based on the data amount at the time when the message is issued. Furthermore, a prediction of future traffic often differs from the actual amount.

The unavailability of the actual traffic amount hinders an optimal bandwidth allocation. {\color{\mycolor} Conventional DBA algorithms} try to solve the problem by estimating the future traffic amount with statistical or machine learning methods. Nevertheless, the static optimization relying on the such estimates may demonstrate high variance in its performance, as the estimation is not an easy task in general even with recent machine learning techniques. Furthermore, such estimation will be more complex when considering the high user mobility in access networks. Hence, it seems desirable to use an allocation algorithm that gradually adjusts its allocation strategy based on the past traffic patterns to obtain an optimal allocation robust against fluctuating traffic over time, instead of changing an allocation at every time cycle to make it the exact optimum for an estimated traffic amount. In this sense, we want to design an allocation algorithm that is numb to traffic changes over a short time span.

Our proposal is to compute the window size $\varx_t$ from the actual traffic from $C_0$ to $C_{t-1}$ by an Online Convex Optimization (OCO) algorithm, instead of solving the static optimization problem with the estimated amount $\hat{\varb}_t$. Since the exact amount of queued data at the previous cycle $C_{t-1}$ is observed after the window allocation $\varx_{t-1}$, our algorithm can use the actual data $\mathbf{b}_{t-1}$ to decide an allocation for the next cycle $C_t$.

By defining a convex version of the objective, $g_t(\varx_t) \triangleq - f_t(\varx_t)$, the maximization problem of $f_t$ is converted to a minimization problem of a convex function $g_t$ over $\varx_t$ with the same constraint in Eq. \eqref{objective}. We use Projected Gradient Descent (PGD) \cite{shalev2011online} to solve this problem. At $C_{t-1}$, the accurate amounts of queued data are collected at the OLT via REPORT messages (See Figure \ref{fig_cycle}). Thus, the actual objective function at the previous cycle $g_{t-1}$ becomes available (Eq. \eqref{objective} with $\mathbf{b}_{t-1}$). Note that this approach is different from the traditional use of REPORT messages in that ONUs report the actual traffic amount after the allocated window at $C_{t-1}$ starts. PGD computes the next window size $\varx_t$ by 
\begin{align}
    \varx_t \gets \Pi_{X}\left(\varx_{t-1} - \eta_{t-1} \nabla g_{t-1}(\varx_{t-1})\right), 
\end{align}
where $\Pi_{X}(\varx)$ is a projection bringing $\varx_t$ back to the space $X$, and $\eta_{t-1}$ is a diminishing learning step size. Since the allocation for the next round is defined as a function of the previous allocation, the recursive relation can be represented as $\varx_t \leftarrow \mathrm{PGD}(\varx_0, ..., \varx_{t-1})$, which was initially desired.

\subsection{OCO Performance Bound}
We are interested in bounding the average performance of a DBA algorithm over time rather than obtaining the exact optimum allocations at each cycle, since the exact optimum solution is not feasible without an oracle reporting the exact future traffic amount $\varb_t$. The \textit{regret} of an algorithm $A$ quantifies the discrepancy between a sequence of allocations by $A$ and a hindsight optimum allocation (the most robust allocation over all cycles till $C_T$):
\begin{align*}
\operatorname{regret}_{T}(A)=\sup_{\left\{g_{1}, \ldots, g_{T}\right\} \subseteq G}\left\{\sum_{t=1}^{T} g_{t}\left(\mathbf{x}_{t}\right)-\min _{\mathbf{x} \in X} \sum_{t=1}^{T} g_{t}(\mathbf{x})\right\}. 
\end{align*}
Note that the supremum over all possible sequences of objective functions implies that the regret bound considers adversarial scenarios, which is, in our case, a situation where the actual traffic pattern is extremely irregular. 

The gradient descent-based solution (PGD), which we use, is known to have $O(\log T)$-bound for the regret with the convergence rate of $1/{\sqrt{T}}$ for a convex objective function over a convex variable range \cite{shalev2011online}.

\begin{figure}[t]
    \centering
    \includegraphics[width=\columnwidth]{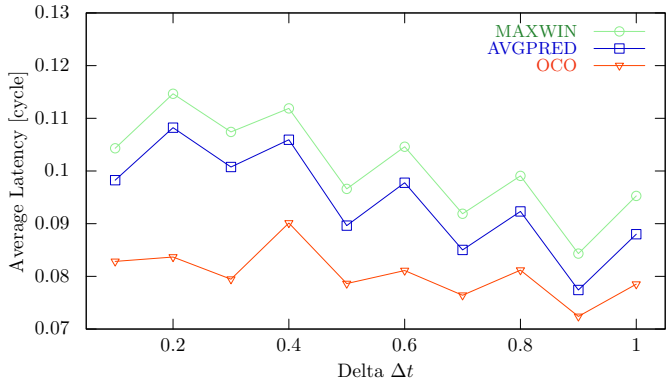} 
    \caption{Average Latency at ONUs in respect to the cycle size $C = 1$.}
    \label{fig_delay}
\end{figure}


\section{Experiment and Discussion}
\subsection{Network Settings}
Simulations are conducted in a network with three slices: $S_0, S_1,$ and $S_2$. The network hosts ten ONUs, and each slice has a different number of ONUs as follows: $\{u_0, u_1, ..., u_4\} \in S_0, \{u_5, u_6, u_7\} \in S_1,$ and $\{u_8, u_9\} \in S_2$. Among the ten ONUs, ONUs $u_0, u_3, u_6,$ and $u_9$ are heavily loaded with a traffic generation function based on the Poisson distribution $\mathrm{Pois}(\lambda = 10)$, while the traffic for the other ONUs is generated based on $\mathrm{Pois}(\lambda = 1)$. Unless otherwise specified, the slice prioritization weight $p_j$ is set to $1.0$ for all slices. It is only changed when evaluating the slice-awareness.

\subsection{Performance Comparison: Typical DBA Algorithms}
The proposed method is compared with typical generic DBA algorithms: MAXWIN and AVGPRED. 
\subsubsection{MAXWIN}
allocates either the predefined maximum window size $m$ or the traffic amount queued in ONUs, which is reported in the previous REPORT message. Thus, this algorithm is quite responsive to the demand from each ONU. While both $m = 0.2$ and 0.4 are tested in the simulations, only the result with $m = 0.2$ is discussed below, since their performances are similar. 

\subsubsection{AVGPRED}
assigns the average of the actual traffic queued at each ONU in past cycles. When computing the average at round $C_t$, the algorithm uses the past traffic amounts from $C_{t-1}$ to $C_{t-h}$, where $h$ is a given time horizon. The performance is shown with a horizon $h$ that provided the best result among different $h$'s ($h \in \{10, 100, 1000\}$). This algorithm stays optimum for every static problem defined with a predicated traffic amount. Note that an allocation vector can be projected back to the solution space to make it a feasible solution, preserving the relative allocation ratio among ONUs.

\subsection{Experimental Results}
We discuss our simulation results from the three aspects desired for a DBA algorithm; namely, demand-responsiveness, slice-awareness, and allocation fairness. In addition, an interpretation of the OCO performance guarantee in our problem is briefly summarized at the end.

\subsubsection{Demand-responsiveness}
Figure \ref{fig_delay} illustrates the average latency versus the delta value ($\Delta t$) depicted in Figure \ref{fig_cycle}, during which additional traffic unknown to a DBA arrives. This value implies the amount of traffic that was not reported to the OLT and can be added to the queues of ONUs before the next round. The y-axis is the average latency of each traffic unit from 50 simulation runs. $\Delta t$ and the average latency are represented with respect to the cycle size $C = 1$.

\begin{figure}[t]
    \centering
    \includegraphics[width=0.85\columnwidth]{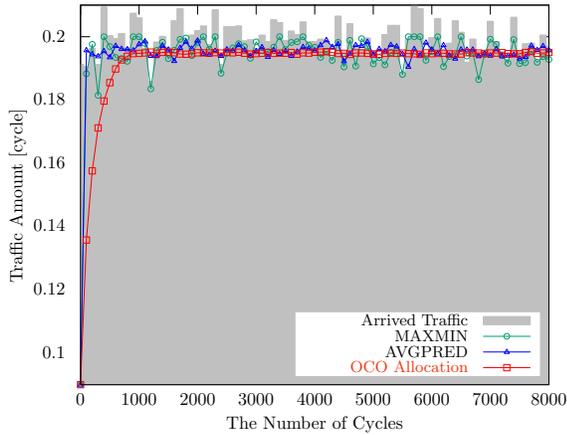} %
    \caption{Transition of the Allocation to ONU $u_0$: The proposed DBA algorithm adjusts its allocation over time to provide a robust bandwidth allocation, considering the past traffic pattern.}
    \label{fig_trans}
\end{figure}
\begin{figure}[t]
    \centering
    \includegraphics[width=0.8\columnwidth]{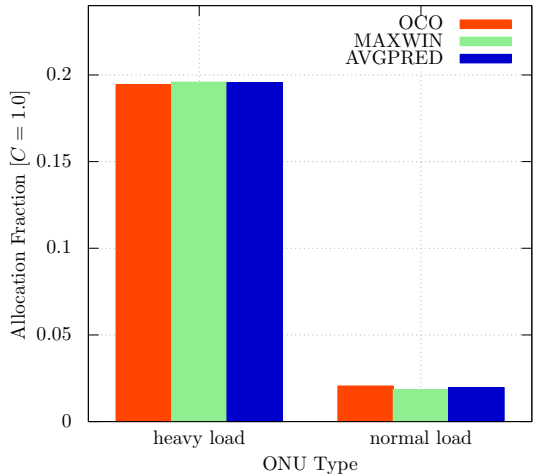} %
    \caption{Allocation Depending on the Traffic Load at $C_{10000}$: A heavily-loaded ONU receives more bandwidth.}
    \label{fig_window_p}
\end{figure}


The proposed OCO-based algorithm realizes the lowest queuing delay, on average, among the three methods. This result indicates that our solution reduces the average latency even though it does not provide the optimum for every single static optimization. This is because the OCO-based approach gradually adjusts its solution to an allocation that is robust against underlying fluctuations of traffic.

The general decreasing trend of MAXWIN and AVGPRED in the average latency could be explained by excessive adjustment towards traffic amount in each time cycle. As $\Delta t$ increases, the total amount of unreported traffic becomes more random. Therefore, the negative impact of an unmatched bandwidth allocation will be smaller in scale \textit{on average}. In contrast, the impact stays notable when $\Delta t$ is smaller. It is worth mentioning that this behavior does not imply the usefulness of MAXWIN and AVGPRED with a larger $\Delta t$. This is because it is reasonable to assume the duration between a GATE message and the previous REPORT message ($\Delta t$) is within one time cycle $C$, and our OCO approach performs better in the range.

Figure \ref{fig_trans} shows the learning process of the allocation to ONU $u_0$ by the proposed algorithm along with the observed actual traffic pattern (grey bars) and the allocations of the other algorithms. The result indicates that the OCO-based algorithm quickly adjusts its allocation to the traffic pattern at earlier cycles and continues fine-tuning at later cycles. The result also indicates the convergence of our OCO-based allocation over time.

{\color{\mycolor}
Figure \ref{fig_window_p} illustrates the converged allocation to ONU $u_0$ at $C_{10000}$ by each method. It indicates that the proposed algorithm allocates slightly more resources to normally loaded ONUs. While the difference in the allocation among the three methods is quite small (on a scale of $10^{-3}$), the results discussed above show that the small difference in allocation succeeds in reducing the average delay. Since the heavily loaded ONUs aggressively request the resources, the other algorithms, which are relatively myopic, tend to allocate more resources towards the ONUs with heavier loads. However, the overallocation could be alleviated in the proposed algorithm due to its gradual allocation updates towards a robust solution.
}

\subsubsection{Slice-awareness}
Table \ref{sa_table} indicates the difference in the allocated window sizes depending on the difference among network slices with different delay-sensitivity. The slice prioritization weight $p_j$ was adjusted to represent the specific delay requirements. In particular, assuming that slice $S_3$ hosts a delay-sensitive network service, the weight $p_3$ of slice $S_3$ was set to 1.2, while the weights for the other slices are 1.0. It is observed that the ONUs on $S_3$ receive more bandwidth. For example, the heavily-loaded ONU $u_9$ on $S_3$ is allocated $0.2642$ cycles at round $C_{10000}$, while the heavily-loaded ONUs on $S_1$ and $S_2$ receive $0.1657$ cycles. This result empirically verifies the possibility to accommodate different types of network slices with diverse requirements through the tuning of the parameter $p_j$.

\subsubsection{Fairness}
{\color{\mycolor}
Table \ref{var_latency} summarizes the average latency in representative ONUs ($u_0, u_1, u_5,$ and $u_6$) and the standard deviation $\sigma_U$ of the delays across all ONUs in $U$. The standard deviation of the proposed OCO-based algorithm is relatively higher than the other methods. While this implies the prioritization of some ONUs, the standard deviation stays in the same order ($10^{-3}$) as the other algorithms. This fact empirically shows that the proposed algorithm maintains the fairness in allocation in terms of the latency, even though it allocates more resources to some ONUs depending on the weight metrics. In other words, the algorithm fairly allocates the resources so that the latency levels stay within a certain range across all ONUs.
}

\begin{table}[t]
    \centering
    \caption{Allocated window sizes with different slice prioritization weights: The slice weight $p_j$ allows delay-sensitive slices to be allocated larger window sizes. (Unit: cycle)\label{sa_table}}
    {\renewcommand\arraystretch{1.4}
    \begin{tabular}{c||c|c}
    \hline
    & Latency-sensitive Slice & Standard Slice\\ 
    & $(p_j = 1.2)$ & $(p_j = 1.0)$\\ 
    \hline
    Heavily-loaded ONU & 0.2642 & 0.1657\\
    \hline    
    Normally-loaded ONU & 0.0569 & 0.0169\\ 
    \hline
    \end{tabular}
    }
    \label{samplerecv}
\end{table}

\begin{table}[t]
    \centering
    \caption{Average Latency at Each ONU and Standard Deviation $\sigma_U$ of the Delays in All ONUs in $U$. (Unit: cycle) \label{var_latency}}
    {\renewcommand\arraystretch{1.4}
    \begin{tabular}{c||c|c|c|c||c}
    \hline
    & ONU 0 & ONU 1& ONU 5 & ONU 6 & $\sigma_U$\\ 
    \hline
    OCO & 0.0351 & 0.0172 & 0.0169 & 0.0352 & 0.0094\\
    \hline
    MAXWIN & 0.0230 & 0.0304 & 0.0302 & 0.0231 & 0.0039\\ 
    \hline
    AVGPRED & 0.0231 & 0.0306 & 0.0310 & 0.0230 & 0.0039\\ 
    \hline
    \end{tabular}
    }
    \label{samplerecv}
\end{table}


\section{Conclusion}
This paper proposes an Online Convex Optimization (OCO) based solution to the Dynamic Bandwidth Allocation (DBA) problem. Our algorithm is aimed at realizing an bandwidth allocation that is slice-aware, demand-responsive, and fair among ONUs in PON access networks, formulating the DBA problem based on the concept of proportional fairness with appropriate weight parameters. The use of the OCO scheme enables our allocation solution to be robust against fluctuations of traffic, which eventually results in the reduction of the average latency over time. The simulation results indicate that the proposed solution mitigates the latency compared to other typical allocation approaches. Furthermore, the results infer the effectiveness of a robust allocation over time for access networks with dynamic traffic patterns, in contrast to the optimum solutions at static optimizations formulated with an estimated traffic pattern.

\section*{Acknowledgement}
This work was supported in part by the National Science Foundation under Grant No. CNS-2008856.

\bibliographystyle{IEEEtran}
\bibliography{ponref}

\end{document}